\documentclass[11pt,twoside]{acta}
\usepackage{epsfig,float,floatflt}
\SetPages{000}{000}
\SetVol{00}{2000}
\usepackage{graphics}

\begin{document}

\begin{Titlepage}
  \Title{Petersen diagram for RRd stars in the Magellanic Clouds}
  \Author{ B.~L.~P~o~p~i~e~l~s~k~i~$^1$,
  	   W.~A.~D~z~i~e~m~b~o~w~s~k~i~$^{1,2}$ \and 
	   S.~C~a~s~s~i~s~i~$^3$}
{$^1$ Warsaw University Observatory, Al.Ujazdowskie~4,~00-478~Warsaw, Poland\\
 $^2$ Nicolaus Copernicus Astronomical Center, ul.Bartycka~18, 00-716
  	Warsaw, Poland\\ 
 $^3$ Osservatorio Astronomico Collurania, I-64100, Teramo, Italy\\
e-mail: popielsk@astrouw.edu.pl, wd@astrouw.edu.pl, cassisi@astrte.te.astro.it}

\vspace*{9pt}
\Received{2000}
\end{Titlepage}

\Abstract{RRd stars from the Magellanic Clouds form a well-defined band in
the Petersen diagram. We explain this observed band with our evolutionary
and pulsation calculations with assumed metallicity [Fe/H]$=(-2,-1.3)$.
Vast majority of RRd stars from LMC is confined to a narrower range of
$(-1.7,-1.3)$. The width of the band, at specified fundamental mode period,
may be explained by mass spread at given metallicity.
The shape of the band reflects the path of RRd stars
within the RR~Lyrae instability strip. We regard the success in explaining 
the Petersen diagram a support for our evolutionary models,
which yield, mean absolute magnitude in the mid of the instability strip,
$\left<M_V\right>$, in the range $0.4$ to $0.65$ mag implying distance
modulus to LMC of $18.4$ mag.}{stars: variable, stars:
oscillations, stars: RR~Lyrae, stars: double-mode pulsations, galaxies: 
Magellanic Clouds, stars: abundances}

%**************************************
\section{Introduction}
%**************************************
RR~Lyrae stars are objects of great importance for whole astrophysics.
Double-mode RR~Lyrae stars (RRd), constitute relatively rare
subtype. Nonetheless RRd stars attract considerable attention because they 
provide us an additional precise observable, which is the second period (see
review of Kov{\'a}cs (2000a)).

Usefulness of double-mode pulsators was first realized by Petersen~(1973).
Following his idea, the period data for double-mode pulsators are commonly
represented in diagrams, called now {\em Petersen diagrams}, in which the
period ratio, ${\cal R}=P_1/P_0$, is plotted against the fundamental mode
period, $P_0$. Published examples of Petersen diagrams for RRd stars may 
be found in the following papers: Nemec~\etal (1985a), Clement~\etal 
(1986), Walker~\etal (1994), Alcock~\etal (1997,2000b), Beaulieu~\etal (1997).
The Oosterhoff dichotomy is manifested in Petersen diagrams for RRd
stars. Namely, RRd variables from the Oosterhoff~I globular clusters occur 
in systematically shorter period range than those from the Oosterhoff~II 
(Smith 1995).

Petersen diagram in its first application was used to derive mass and
radius estimates for Cepheids (Petersen 1978).
The result was a huge discrepancy, known as the double-mode Cepheid mass
problem, between the masses derived in this way
and the evolutionary masses. The problem disappeared once the OPAL opacities
became available (Moskalik~\etal 1992).

First comparison of the observed and theoretical Petersen diagrams for
RR~Lyrae stars was made by Cox~\etal (1980). No large discrepancy
between observations and models was found.
The new opacities were first used to calculate period ratios for RRd
models by Cox (1991), later by Kov\'{a}cs~\etal (1991,1992) and 
Bono~\etal (1996). Kov\'{a}cs~\& Walker (1999) used data on RRd stars from 
globular clusters to derive luminosities of RR~Lyrae stars yielding support for
a brighter luminosity scale.

A large number of data on RRd stars became available as a byproduct of 
micro-lensing surveys. Most of currently known RRd stars were observed by
MACHO collaboration (Alcock~\etal 1997,2000b).
Figure~1. shows Petersen diagrams for all known RRd stars. The
objects belonging to different stellar systems, galaxies and
globular clusters, are shown in separate panels with big dots.
We see that RRd stars form quite a narrow curved band in the
Petersen diagram. 
The objects from LMC are spread all over the period range, which
is $0.46-0.6$ d. Short period end in LMC is significantly
more populated. We don't see such a concentration for the SMC
objects which are more-or-less uniformly distributed, however in view of
much scarcer data this conclusion must be regarded preliminary. 
The data for the Galaxy and dwarf galaxies Draco and
Sculptor are too sparse to conclude anything about properties of RRd
stars in these systems. The objects from globular clusters are
localized in small parts of the band.

The goal of our paper is to explain the properties of the RRd band as
defined by the objects from LMC and SMC. From this we expect to
learn something about RR~Lyrae star properties, particularly their 
luminosities, as well as about early history of star formation in the 
Magellanic Clouds. Furthermore, the abundant data on RRd stars may yield
us a useful constraint regarding the origin of double-mode pulsation. Although 
the double-mode pulsation has been successfully modeled (e.g. by Feuchtinger (1998)), our understanding of its causes is not yet satisfactory. 

Application of Petersen diagrams as a probe of stellar properties is
explained in Section~2. In Sections~3. and~4. we provide some details
regarding our model and pulsation calculations, respectively.
Interpretation of the RRd band based on model
calculation are presented in Section~5. In Section~6. we 
discuss absolute magnitudes of RRd stars. Uncertainties of the theoretical 
Petersen diagram are analyzed in Section~7.

\begin{figure}[hbt]
 \centering
 \epsfig{file=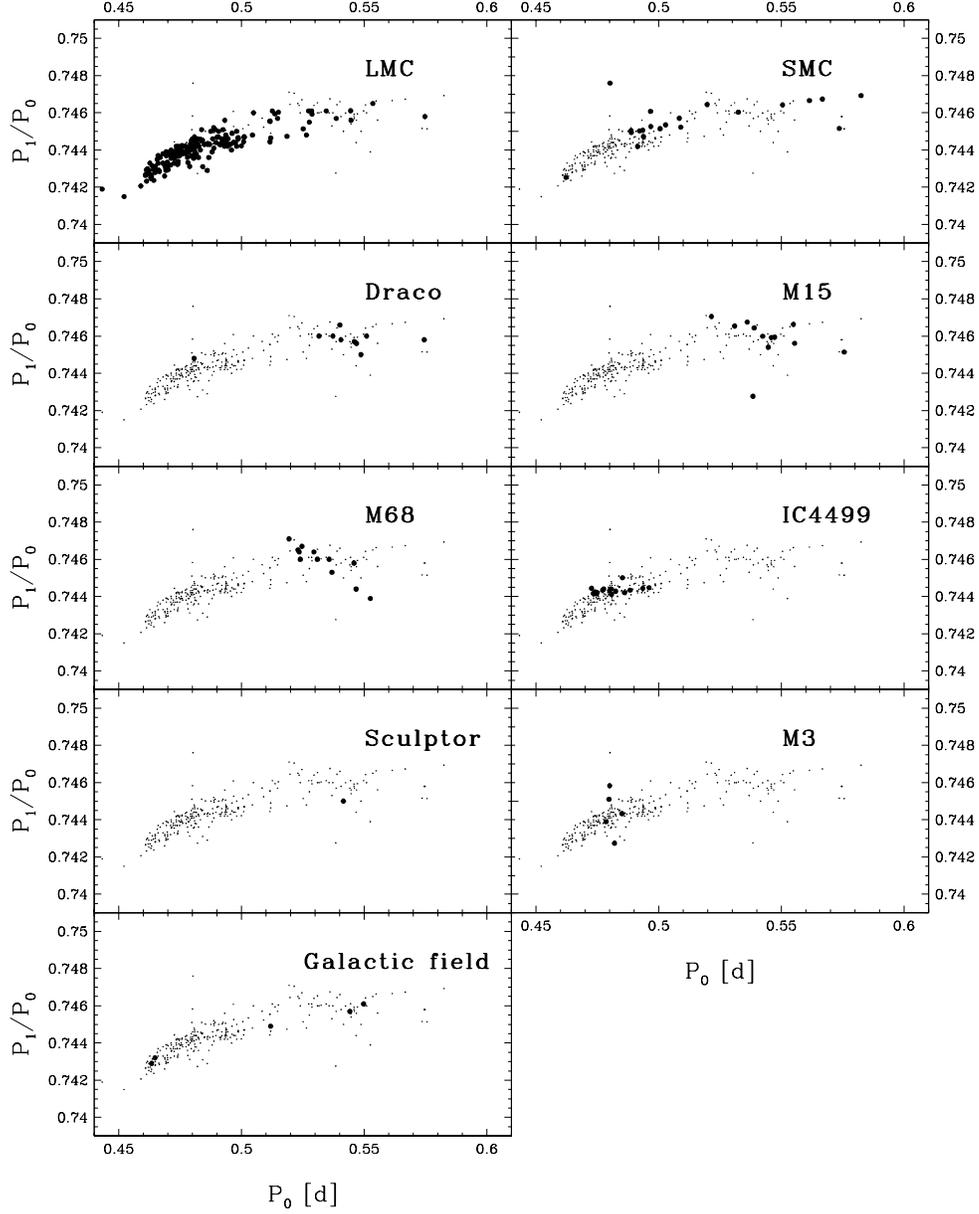,height=1.27\linewidth,bbllx=21,bblly=40,bburx=572,
bbury=740}
 \caption[]{{\em Petersen diagrams} for RRd stars in various stellar
 systems. Data taken from (number of objects in parentheses):
{\bf LMC}~(181) -- Alcock~\etal (1997,~2000b),
{\bf SMC}~(26) -- OGLE private communication (unpublished),
{\bf Draco}~(10) -- Nemec (1985a),
{\bf IC4499}~(16) -- Clement~\etal (1986), Walker~\&~Nemec (1996),
{\bf M3}~(5) -- Corwin~\etal (1999),
{\bf M15}~(12) -- Nemec (1985b), Jurcsik~\&~Barlai (1990), Purdue~\etal (1995),
{\bf M68}~(12) -- Walker (1994),
{\bf Sculptor}~(1) -- Ka{\l}u{\.z}ny~\etal (1995),
{\bf Galactic field}~(5) -- Garcia-Melendo~\&~Clement (1997),
 Clementini~\etal (2000).}
 \label{f:obs}
\end{figure}

%**************************************
\section{Petersen diagram astrophysics}
%**************************************
\label{s:pda}
One needs six parameters to calculate envelope structure and
radial mode frequencies. These are, for example: mass, luminosity, effective
temperature, two parameters for chemical composition (i.e. $Y$
and $Z$) and the mixing-length theory parameter ($\alpha$). 
We fixed the value of $X$ at $0.76$ because possible small variations about 
such a value have little effect on stellar properties. We also fixed the value 
of $\alpha$ in our main surveys,
however in Section~7. we discuss uncertainties connected with the choice of
$\alpha$. Also in Section~7. we discuss effects of choosing heavy
element composition different than the solar mix (Grevesse~\& Noels 1993)
adopted by us as a standard in pulsation calculations. We use this standard
to translate $Z$ to [Fe/H]. This quantity is more customary than [M/H].

Figure~2. shows how the four remaining
parameters affect position of the RRd model in the Petersen diagram. The
reference model may be regarded typical for RRd stars in LMC. We may see
that the parameter crucial for the value of period ratio is $Z$, which is
not a new observation. Next important parameter is luminosity, $L$, and mass, 
$M$. However, $M$ and $L$ are strongly constrained by $Z$, as we will see in 
next sections. Thus, we may regard the value of period ratio as a probe of 
metallicity. If we assume that the mass of RR~Lyrae star is determined by $Z$, 
which is only approximately correct, then with the help of stellar evolution
calculations, which yield $L(M,T_{\rm eff},Z)$ (in fact, two of them
because of the shape of the track), we obtain a one-to-one correspondence 
between a trajectory in the Petersen diagram and the $T_{\rm eff}(Z)$ 
dependence for RRd stars. In reality, we do not have unique $M(Z)$ dependence, 
thus we have a band rather than a single trajectory. Still, the shape of the 
band must reflect primarily the $T_{\rm eff}(Z)$, or equivalently, 
$L(T_{\rm eff})$ relation for RRd stars.

\begin{figure}[hbt]
 \centering
 \epsfig{file=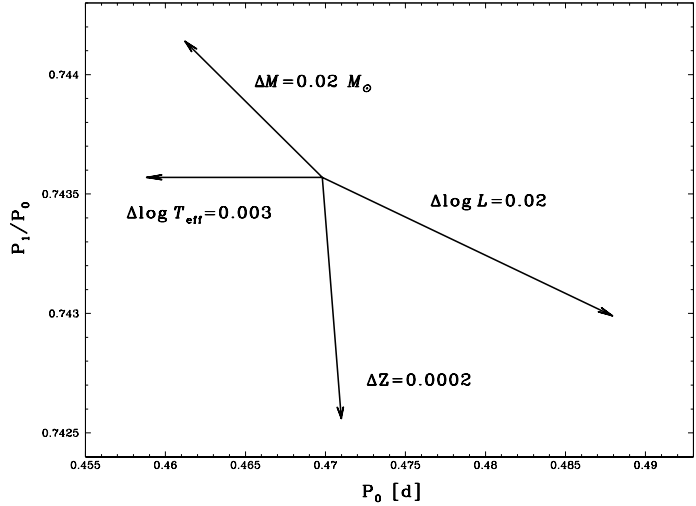,height=.72\linewidth,angle=270,bbllx=41,bblly=12,
	 bburx=583, bbury=741}
 \caption[]{The effect of the envelope parameters on the model position
  in the Petersen diagram. The central model is characterized by the following
  parameters: $X=0.76$, $Z=0.0007$, $\log{T_{\rm eff}}=3.842$,
 $\log{L/L_\odot}=1.705$ and $M = 0.69\:M_\odot$.}
 \label{f:local}
\end{figure}

%**************************************
\section{Evolutionary models}
%**************************************
\label{s:em}
All the evolutionary tracks for He-burning stellar models used
in the present analysis, have been obtained by means of the FRANEC
evolutionary code (Stra\-nie\-ro~\& Chieffi 1989, Cassisi~\& Salaris 1997,
Castellani~\etal 1999), by adopting canonical semiconvection
for the treatment of mixing during the central He-burning phase
(Horizontal Branch, hereinafter HB).

The OPAL radiative opacities (Iglesias~\& Rogers 1996) have been
adopted for temperatures higher than $10,000$ K, while for lower
temperatures we have used the molecular opacity tables provided
by Alexander~\& Ferguson (1994). This choice allows us to have a
smooth match between the two different opacity sets.
Both high and low-temperature opacities have been computed by
assuming a solar scaled heavy element distribution (Grevesse
1991). As far as it concerns the equation of state, we adopted
the Straniero (1988) EOS supplemented at lower temperatures with
a Saha EOS. The outer boundary conditions have been fixed according to the 
$T(\tau)$ relation given by Krishna-Swamy (1966). Concerning the treatment 
of the superadiabatic layers the mixing-length calibration
provided by Salaris~\& Cassisi (1996) has been adopted.
The other physical inputs are the same as in Cassisi~\& Salaris (1997).

We have computed evolutionary sequences of models for several metallicities, 
but by adopting in all cases an initial helium abundance equal to $Y=0.23$.
All the HB models correspond to a Red Giant Branch progenitor with mass
equal to $0.8\;M_\odot$. In fact, the Zero Age HB structures have been
constructed, for each fixed metallicity, by using the helium core mass 
and the envelope chemical abundance profile suitable for a RGB progenitor 
with this mass.

%**************************************
\section{Pulsations}
%**************************************
\label{s:p}

Our next step was to map evolutionary tracks into Petersen diagram.
The envelope models were calculated for surface parameters taken from the
evolutionary tracks. The linear non-adiabatic periods were calculated with a
standard pulsation code (Dziembowski 1977). The envelopes were calculated 
with slightly different physics than the evolutionary models. OPAL
equation of state was adopted in whole envelope. Although there are very small
differences between HB tracks computed with the OPAL compared to Straniero 
(1988) equation of state, the differences in resulting periods could be 
significant. The depth of the envelope and the spatial resolution were 
determined by the accuracy requirement of period ratios on the level of 
$2\times 10^{-4}$. Convection was treated with standard MLT formalism. In
pulsation calculation we ignored the Lagrangian perturbation of the convective 
flux.

We restricted our attention to the central part of the pulsational instability
strip, extending from $\log{T_{\rm eff}}=3.815$ ($6531$ K) to 3.855 ($7161$
K). In this region both fundamental mode and first overtone pulsations are
unstable. Feuchtinger (1999) in his numerical simulations found either-or
behavior in somewhat narrower temperature range. In another paper Feuchtinger 
(1998) found a sustained double-mode pulsation at $T_{\rm eff}=6820$ K.

The segments of the evolutionary tracks from the selected temperature
ranges were mapped into the Petersen diagram. Examples are shown in
Figure~3.

%**************************************
\section{Properties of the RRd band}
%**************************************
\label{s:prb}

%**************************************
\subsection{Constraints on mass from evolutionary models}
%**************************************
\label{s:ec}

Each of our evolutionary tracks is characterized by two parameters, i.e.
mass and heavy element abundance of the model ($M$, $Z$). Our goal here is to
determine range of $M$ for specified $Z$, corresponding to the selected 
temperature range (Section~4.). The track is considered allowed if the star
spends sufficiently long time, $t_s$, in this range. Figure~4. shows star 
lifetimes for models with $Z=0.0002$ and a range of masses. In Figure~5. we 
show the total time spent in the selected temperature range ($t_s$) as 
function of $M$ and $Z$. We see that the time depends strongly on $M$. In the
cases of two stays in the selected temperature range, like for $M=0.74\;
M_\odot$ shown in Figure~4., $t_s$ is the sum of two time intervals.
For each value of $Z$ we see sharp maxima of $t_s$ ($t_s^{max}$).
The values of $M$ corresponding to the maximum may be described by the 
following mass-metallicity relation:
\begin{equation}
  M/M_\odot=0.709-0.128\; (\mbox{[Fe/H]} + 1.6).
\end{equation}
We regarded the $M$ value as allowed if $t_s\ge 0.2\; t_s^{max}$. The range
of allowed masses for each $Z$ is given in Table~1.

\begin{table}
\centering
\begin{tabular}{|l|ll|}
\hline\hline
Z&$M_{min}$&$M_{max}$\\
\hline\hline
0.0015&0.62&0.675\\
0.001&0.64&0.695\\
0.0005&0.665&0.715\\
0.0003&0.7&0.751\\
0.0002&0.735&0.785\\
0.0001&0.79&0.855\\
\hline\hline
\end{tabular}
\caption[]{The allowed masses of RR~Lyrae stars in the selected
effective temperature range based on the evolutionary tracks.}
\label{t:t1}
\end{table}

%**************************************
\subsection{Constraints on metallicity from Petersen diagram}
%**************************************

We have seen in Figure~2. that two parameters most influencing the 
period ratio value are $M$ and $Z$. In the previous section we showed that the
mass range is significantly limited for given metallicity. Hence, in
practice the period ratio determines $Z$. We can
thus determine its ranges by mapping tracks into the Petersen diagram.
Figure~3. shows tracks for 3 metallicities and corresponding
masses. Mapped tracks allow us to conclude that observed RRd star
metallicities range from $0.0002$ to $0.001$. The resulting RRd ``island'' in
$M-Z$ plane is shown in Figure~6.

We are now in the position to discuss metallicities of RRd stars in the
Magellanic Clouds. With the help of Figure~1. and 3.b we may conclude
that most of RRd stars in LMC have $Z$ values in the range of
$(0.0004,0.001)$, which translates to [Fe/H]$=(-1.7,-1.3)$. 
How this value is compared with determinations by others? Kov{\'a}cs (2000a)
quotes the [Fe/H] spread of $(-1.9,-1.3)$ for all RRd stars in LMC, what is
essentially equivalent to an early estimate of Popielski~\& Dziembowski 
(2000). Clementini~\etal (2000) find a wider range of [Fe/H] from their 
spectroscopic data for RR~Lyrae stars. They quote the $(-2.28,-1.09)$ 
range. At this point we wish to remind the reader that our [Fe/H] values
are inferred assuming the same heavy element abundances as in the
sun. Therefore, caution is needed when comparing those values with the
spectroscopic ones.

Dolphin (2000) finds evidence for a strong star formation episode at
[Fe/H]$=-1.63\pm 0.10$ in LMC. The [Fe/H] values for LMC globular clusters
containing RR~Lyrae stars are in the range $(-2.11,-1.71)$
(Johnson~\etal 1999). We see in Figure~1. that RRd stars from M3 and IC4499
occur in the region of maximum concentration of RRd stars from LMC. The
values of metallicity are respectively $-1.57$ and $-1.5$ (Smith 1995), which
are close to the center of the metallicity range inferred by us. On the other
hand, RRd stars from M15 and M68 occur at the high period end of the RRd
band. The metallicities of these two clusters are $-2.15$ and $-2.09$ (same
source), lower by some $0.2$ than those inferred by us. As we discuss in
Section~7., this discrepancy may be explained by too low value of $\alpha$
adopted in our pulsation calculation, but also could be due to uncertainty
of [Fe/H], which exceeds $0.15$ (see for instance Rutledge~\etal (1997)).

\begin{figure}[hbt]
 \centering
 \epsfig{file=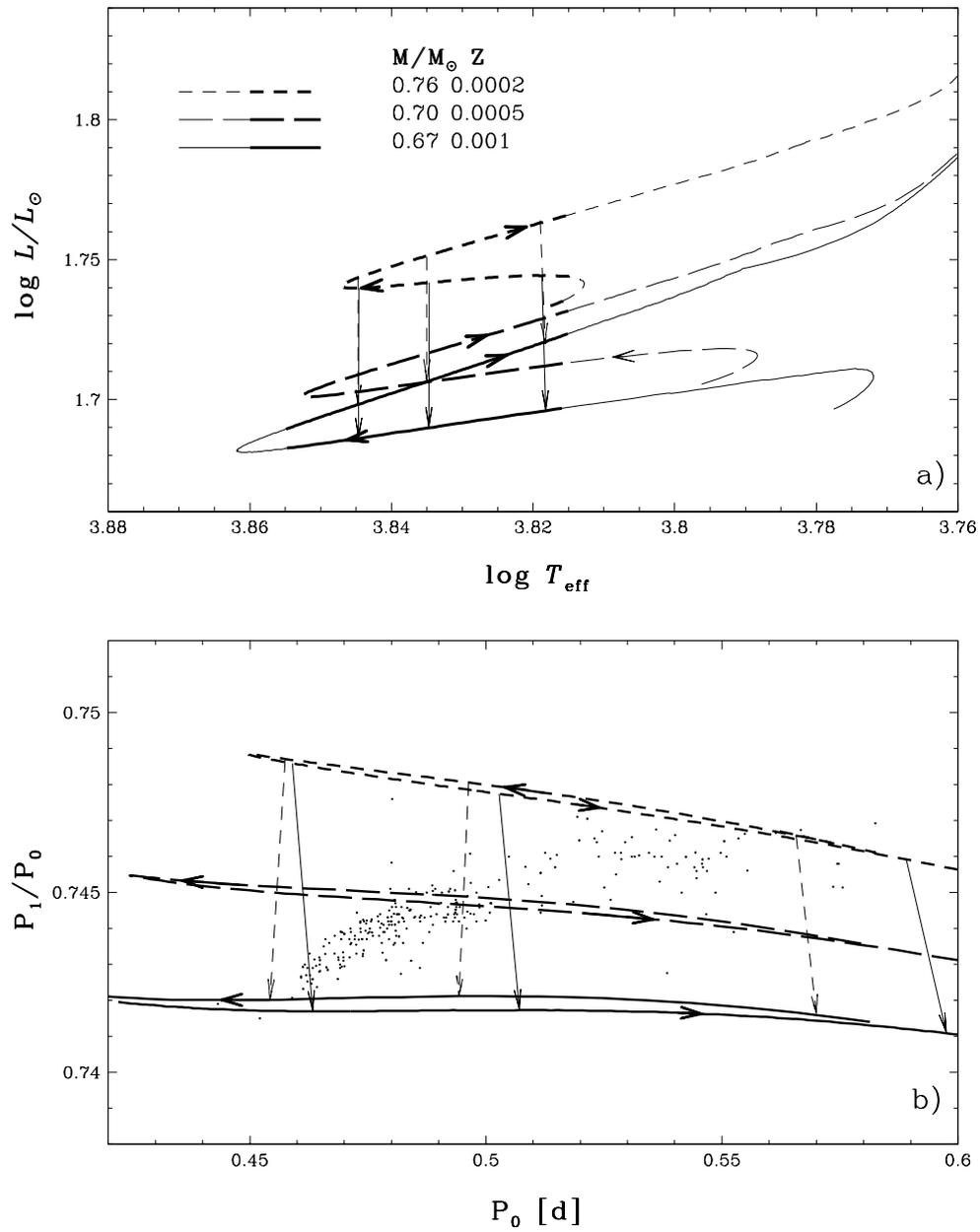,height=1.312\linewidth,bbllx=21,bblly=40,bburx=572,
	 bbury=740}
 \caption[]{Selected evolutionary tracks in: {\bf a)}~{\em the 
Hertzsprung-Russel diagram}, {\bf b)}~Petersen diagram. In {\bf a)}~we use 
thick lines to denote parts of the tracks used in pulsation calculations. 
Observed RRd stars are shown as dots in {\bf b)}. Vertical arrows connect 
models with the same effective temperature, from the left: 
$\log{T_{\rm eff}=3.845}$, $3.835$ and $3.818$. Solid and dashed arrows 
correspond to the lower and upper branch of the tracks, respectively.}
 \label{f:tory}
\end{figure} 

\begin{figure}[hbt]
 \centering
 \epsfig{file=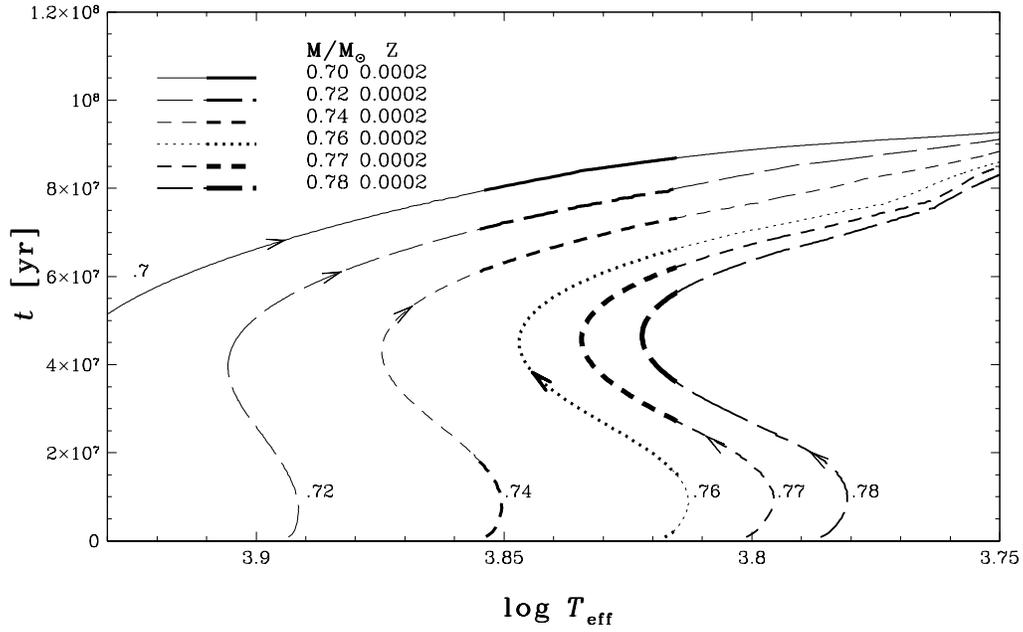,height=.65\linewidth,bbllx=21,bblly=405,bburx=572,
	 bbury=740}
 \caption[]{Stellar evolution lifetime from ZAHB versus temperature for
 different mass RR~Lyrae tracks. Lines are thicker within the
 instability strip.} 
 \label{f:rrtime}
\end{figure}

\begin{figure}[hbt]
 \centering
 \epsfig{file=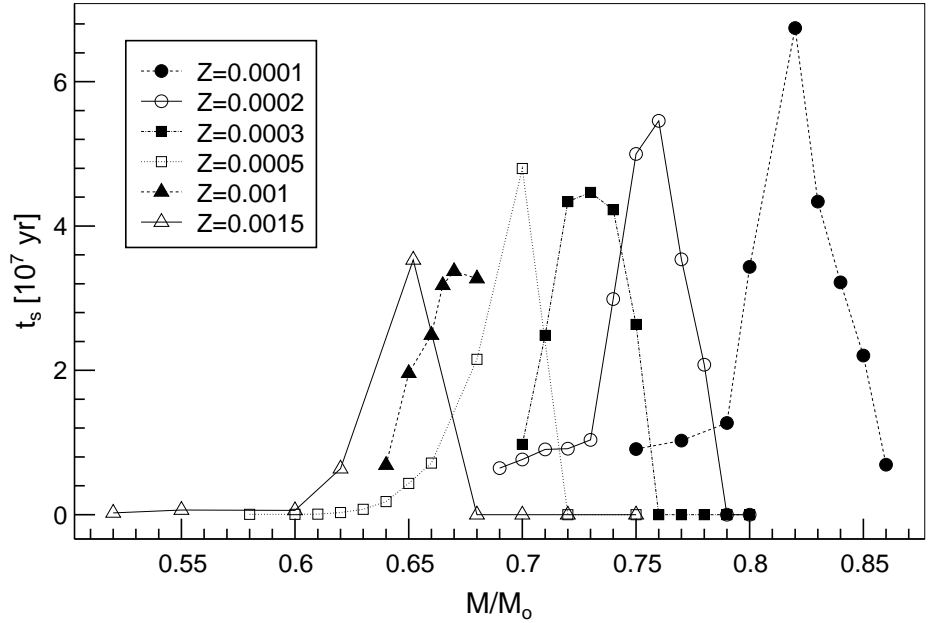,height=.65\linewidth,bbllx=24,bblly=465,bburx=538,
	 bbury=810}
 \caption[]{Model evolution lifetime in the selected temperature range for
 tracks of different mass and $Z$.} 
 \label{f:rrtime}
\end{figure}

\begin{figure}[hbt]
 \centering
 \epsfig{file=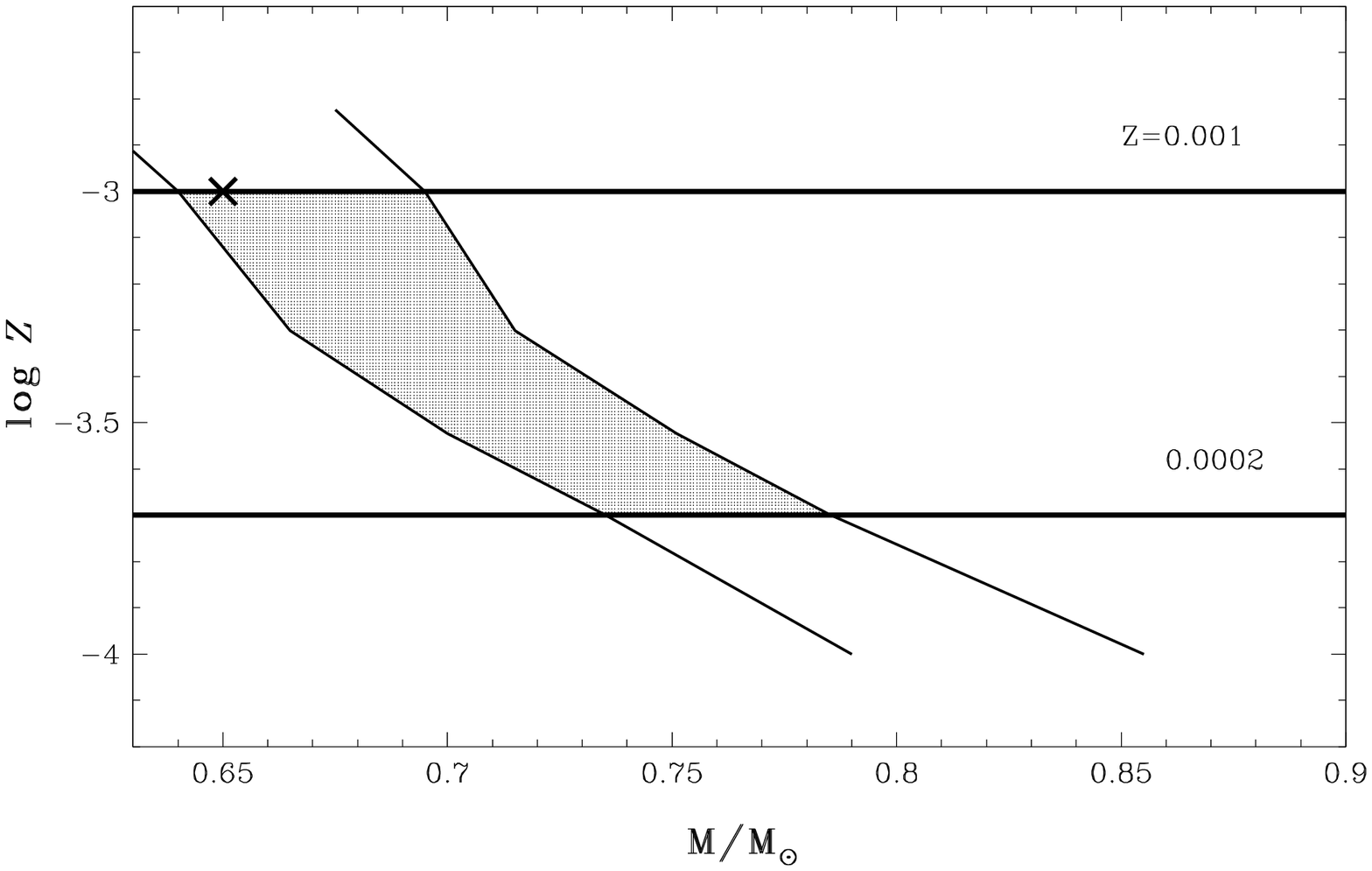,height=.65\linewidth,bbllx=21,bblly=405,bburx=572,
	 bbury=740}
 \caption[]{Allowed domain for RRd stars in the $M-Z$ plane. 
Two sloping $M(Z)$ lines are determined by considering time spent in the
selected temperature range. Two horizontal lines follow from the observational
Petersen diagram. Nonlinear double-mode model (Feuchtinger 1998) is shown
with a cross.}
\label{f:masy}
\end{figure}

%**************************************
\subsection{Width and shape of the RRd band}
%**************************************

The width of the RRd band at given $P_0$
may be explained by the mass spread for specified metallicity.
The estimate of the corresponding spread in period ratio, $\Delta {\cal R}$, 
is provided in Table~2. We may see that it ranges from $0.002$ to $0.004$, 
which is even more than the width of the observed band (see Figure~3.b). In 
the same figure we may see that the branch ambiguity is a relatively small 
contributor to the spread in the period ratio. 

\begin{table}
\centering
\begin{tabular}{|l|l|l|l|}
\hline\hline
Z&$\Delta M$&$\left(\frac{\partial
{\cal R}}{\partial M}\right)_{P_0}$&$\Delta{\cal R}$\\
\hline\hline
0.001&0.055&0.0667&0.0037\\
0.0005&0.050&0.0500&0.0025\\
0.0003&0.051&0.0467&0.0024\\
0.0002&0.050&0.0428&0.0021\\
\hline\hline
\end{tabular}
\caption[]{Mass spread effect on the Petersen diagram band width at
constant $P_0$.}
\label{t:width}
\end{table}

To simplify further discussion of the shape, we adopt from now on a unique
$M(Z)$ relation that is determined by the maxima shown in Figure~5. With this
restriction we may write
\begin{equation}
P_{0,k}=P_{0,k}(Z,T_{\rm eff}),
\end{equation}
\begin{equation}
{\cal R}_{\:\;k}={\cal R}_{\:\;k}(Z,T_{\rm eff}),
\end{equation}
where $k=1,2$ identifies the branch of the track. After eliminating 
$T_{\rm eff}$, we 
get the Petersen relations ${\cal R}_{\;\:k}={\cal R}_{\;\:k}(P_0,Z)$. In 
Figure~3.b we see that for
given $Z$, the separation reflecting the two choices of $k$, is well
within the observed width. Thus, the shape of the RRd branch reflects
the $T_{\rm eff}(Z)$ dependence for RRd stars.

Adopting central values of ${\cal R}$ at given $P_0$, we may invert Eq. 
(2)~and (3) to obtain $T_{\rm eff}(Z)$. In this case, the branch ambiguity
is significant. Using our evolutionary tracks and Eq. (1) we found 
\begin{equation}
 \begin{array}{ll}
  \log{T_{\rm eff}} = 3.8347 + 0.0363\; (\mbox{[Fe/H]} + 1.6) & \mbox{ for the upper branch},\\
  \log{T_{\rm eff}} = 3.8315 + 0.0377\; (\mbox{[Fe/H]} + 1.6) & \mbox{ for the lower branch}.\\
 \end{array}
\end{equation}

Successful nonlinear models of double-mode pulsation should explain these
phenomenological relations. At this stage we cannot say whether it is
metallicity or luminosity that matters, since the two parameters ($T_{\rm
eff}$, [Fe/H]) are correlated.
With the use of evolutionary tracks we can determine the 
corresponding $T_{\rm eff}$ on $L$ relations, which give the locus of RRd 
stars in the HR diagram. The result is
\begin{equation}
\label{e:hr}
 \begin{array}{ll}
  \log{L/L_\odot} = 1.723 - 2.684\; (\log{T_{\rm eff}-3.835)} & \mbox{ for the upper branch},\\
  \log{L/L_\odot} = 1.703 - 2.124\; (\log{T_{\rm eff}-3.835)} & \mbox{ for the lower branch}.\\
 \end{array}
\end{equation}

These two relations are shown in Figure~7. In the same figure we show 
lines separating different behavior of RR~Lyrae stars from observations 
and theory.

\begin{figure}[hbt]
 \centering
 \epsfig{file=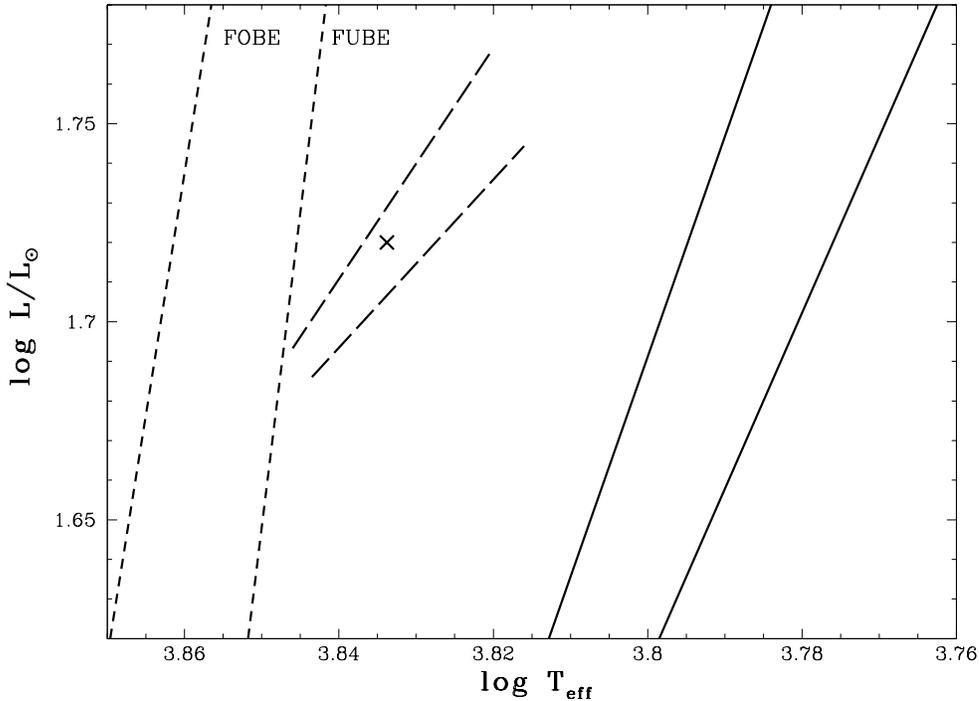,height=\linewidth,angle=270,bbllx=49,bblly=28,bburx=572,
	 bbury=753}
 \caption{Different pulsational behavior or RR~Lyrae in HR diagram. 
 Solid lines encompass observational fundamental mode 
 domain (G{\'e}za Kov{\'a}cs - private communication). The two short-dashed 
 lines, which are from model calculations of Koll{\'a}th~\etal (2000), show
 blue edges for first overtone and fundamental mode pulsators (FOBE and FUBE 
 respectively).
 The two dashed lines show relations given in Eq.~(5). Position of the
 model for which Feuchtinger (1998) found a sustained double-mode pulsation 
 is shown with a cross.}
 \label{f:pasy}
\end{figure}

The RRd localization between RRc and RRab stars was found
observationally by Walker (1994) in his work on RR~Lyrae in M68. It is
supported by recent work of Bakos~\& Jurcsik (2000) for M3.

It is important to notice that inclination, 
$-d\log{(L/L_\odot)}/d\log{T_{\rm eff}}$, of the bimodality strip is
significantly smaller than the inclination of all other characteristic
lines shown in Figure~7. According to current
understanding, the bimodal behavior is caused by specific property of 
convection (see e.g. Koll{\'a}th~\& Buchler (2000)).
Convection also determines the red edge of the instability strip, thus
we would expect a similar inclination of the two lines. That the theory 
fails to explain inclinations of the characteristic lines is not a new 
problem. Indeed, as Koll{\'a}th~\etal (2000) pointed out there is a large 
disagreement between the observational and theoretical lines separating 
first overtone and fundamental mode pulsators. The bimodality strip from 
recent model calculations of Koll{\'a}th~\etal (2000) is drastically 
different than the one determined by us. Apparently, there is still a need 
for further improvement in nonlinear modeling of RR~Lyrae star pulsation.

%**************************************
\section{Double-mode pulsators and absolute magnitudes of RR~Lyrae stars}
%**************************************
\label{s:am}

Kov{\'a}cs~\& Walker (1999) used period and photometric data to determine
the absolute magnitudes of RRd stars from three globular clusters. This was
a novel approach to the important and still vividly debated problem of
absolute magnitudes of RR~Lyrae stars. Different methods yield results, 
that may differ as much as $0.3$ mag. Kov{\'a}cs~\& Walker (1999) found 
RR~Lyrae stars brighter by $0.2\div 0.3$ mag than inferred by means of the
Baade-Wesselink method. Their result was amongst those indicating higher
luminosity of these stars.

In our investigation we make a different use of RRd stars. Instead of 
photometric data, we use evolutionary tracks. Following Kov{\'a}cs~\& Walker 
(1999) we assume that the luminosities of RRd stars are representative 
for the whole population of RR~Lyrae stars. Choosing evolutionary 
track at each $Z$ for the central mass value, we may infer $L$ at the RRd 
temperatures. We recall the [Fe/H] is the customary counterpart of $Z$,
based on Grevesse~\& Noels (1993). With the help of Kurucz (1999) tabular 
data, we derived
\begin{equation}
 \begin{array}{ll}
  \left<M_{V}\right> = 0.452 + 0.162\; (\mbox{[Fe/H]} + 1.6) & \mbox{ for the upper branch},\\
  \left<M_{V}\right> = 0.546 + 0.240\; (\mbox{[Fe/H]} + 1.6) & \mbox{ for the lower branch}.\\
 \end{array}
\end{equation}
The luminosities given in these equations are close
to the high luminosity end of the debated range. 

Somewhat higher luminosities for RR~Lyrae stars were derived by
Carretta~\etal (2000), who relied on the main sequence fitting. He quotes the
following result,
\begin{equation}
 \left<M_V\right>=(0.45\pm0.12) + (0.18\pm0.09)\; (\mbox{[Fe/H]} + 1.6).
\end{equation}
Demarque~\etal (2000) combining their evolutionary tracks with photometric
data on globular clusters obtained
\begin{equation}
 \left<M_V\right>=0.55 + 0.21\; (\mbox{[Fe/H]} + 1.6),
\end{equation}
which is not too different from our estimate. 

There are three distinct methods which lead to the significantly lower 
RR~Lyrae star luminosities.
Gould~\& Popowski (1998) used the statistical parallax method to infer
\begin{equation}
 \left<M_V\right>=0.77\pm0.13 \mbox{ at [Fe/H]}\approx-1.6.
\end{equation}
Udalski~\etal (1999) in their determination of the absolute magnitudes
relied on the distance to LMC, determined by the red-clump
method, and their photometry of RR~Lyrae stars. They found the mean value
of $0.71\pm0.07$.
Smith (1995) quotes two results obtained by means of the Baade-Wesselink
method,
\begin{equation}
\begin{array}{ll}
 \left<M_V\right>=0.76 + 0.16\; (\mbox{[Fe/H]} + 1.6) & \mbox{after (Jones~\etal 1992)},\\
 \left<M_V\right>=0.72 + 0.20\; (\mbox{[Fe/H]} + 1.6) & \mbox{after (Cacciari~\etal 1992)}.\\
\end{array}
\end{equation}
We should stress however, that more recent analysis of the observational
material suggests higher luminosity values
\begin{equation}
\begin{array}{ll}
 \left<M_V\right> = 0.63 + 0.21\; (\mbox{[Fe/H]} + 1.6)& \mbox{after (Fernley 1994)},\\
 \left<M_V\right> = 0.50 + 0.28\; (\mbox{[Fe/H]} + 1.6)& \mbox{after (McNamara 1997)}.\\
\end{array}
\end{equation}
Having determined mean luminosities of RR~Lyrae stars we may provide our
estimates of distance modulus to LMC. In this we must rely on
measurements of visual magnitudes. Mean visual magnitude of RRd stars
from the MACHO data is $19.33$ mag (Alcock~\etal 2000a). From Eq.~(6), 
adopting [Fe/H]=-1.5, we get
average absolute magnitude of $0.52$ mag, which implies
$(m-M)_{\rm LMC}=18.81$ mag. If instead of MACHO we adopt OGLE mean visual
magnitude (Udalski~\etal 1999), we get $(m-M)_{\rm LMC}=18.42$ mag.
The OGLE mean value refers to a sample of various types of RR~Lyrae 
pulsators, but we have no evidence, there is a systematic difference in
luminosities between RRd stars and the whole population of RR~Lyrae stars.
These numbers may be compared with $(m-M)_{\rm LMC}=18.53$ mag, derived by
Kov{\'a}cs (2000a), by means of the method developed by Kov{\'a}cs~\& Walker
(1999). Kov{\'a}cs (2000b) application of a similar method to double-mode
Cepheids from SMC implies, after using distance modulus difference between
SMC and LMC of $0.51$ mag (Udalski~\etal 1999), $(m-M)_{\rm LMC}=18.54$ mag. 
A recent measurement of the LMC distance modulus based on the Red-Clump method
(Romaniello~\etal 2000), as well as other works based on different distance
indicators - referenced therein, supports also a long-distance scale.

%**************************************
\section{Uncertainties}
%**************************************
\label{s:uncert}

We have seen that the Petersen diagram is indeed a powerful tool for
diagnosing evolutionary models of RR~Lyrae stars and for determining
metallicities in stellar systems. The tool, however, requires high
precision in calculated period ratios, ${\cal R}$. A $10^{-4}$ difference in
the period ratio corresponds to 3\% difference in $Z$. Let us recall that
the whole range of period ratios for RRd stars is $0.742\div 0.748$. It is not
difficult to reach the numerical accuracy of $2\times 10^{-4}$ in calculated 
period ratios within the linear non-adiabatic treatment. Typical difference
between non-adiabatic and adiabatic period ratios is $0.002$.

The treatment of convection has some effect on calculated periods.
Undoubtedly, the $\alpha$ effect on ${\cal R}$ 
cannot be ignored. The period ratio values are affected by the difference 
in the envelope structure implied by change in $\alpha$.
In Figure~8. we illustrate the effect of $\alpha$ on the 
interpretation of the Petersen diagram. We compare there model results 
obtained with $\alpha=1$, which was the adopted standard in our pulsation
calculations, with $\alpha=2$, close to the values adopted in evolutionary
track calculations. The choice of $\alpha=2$ in pulsation calculation implies
shift of upper limit of $Z$ from $0.001$ to $0.00125$ ([Fe/H] from 
$-1.3$ to $-1.2$) and the shift of the lower limit of $Z$ -- from 

Details of heavy element composition are another effect of potential 
significance. In our calculations we used the standard solar mixture 
(Grevesse~\& Noels 1993). Kov{\'a}cs~\etal (1992) studied the effect of 
using different mixtures. Specifically, they considered an oxygen enhanced 
mixture. Using tabular data of Kov{\'a}cs~\& Walker (1999) we found that 
with such a mixture, the maximum value of $Z$ for RRd stars should be about 
$0.002$ instead of $0.001$, quoted by us in Section~5.2. However, the 
change in the metallicity parameter, [Fe/H], isn't high and amounts to 
about $-0.15$.

The uncertainty of nonlinear effect on the value of period ratio is the most 
difficult to estimate. The problem was first investigated by Bono~\etal (1996)
and recently revisited by Koll{\'a}th~\& Buchler (2000). These authors found 
that the period ratio shift due to nonlinear effect is less than zero, in
most cases, and may be as large as $-8\times 10^{-4}$. We will use the value 
of $-4\times 10^{-4}$, which corresponds to typical amplitudes of
RR~Lyrae stars. Thus we have to add a $4\times 10^{-4}$ correction to 
infer the linear period ratio value from observations. This
correction translates to 10\% decrease in the inferred value of $Z$
($0.05$ decrease in [Fe/H]). 

\begin{figure}[hbt]
 \centering
 \epsfig{file=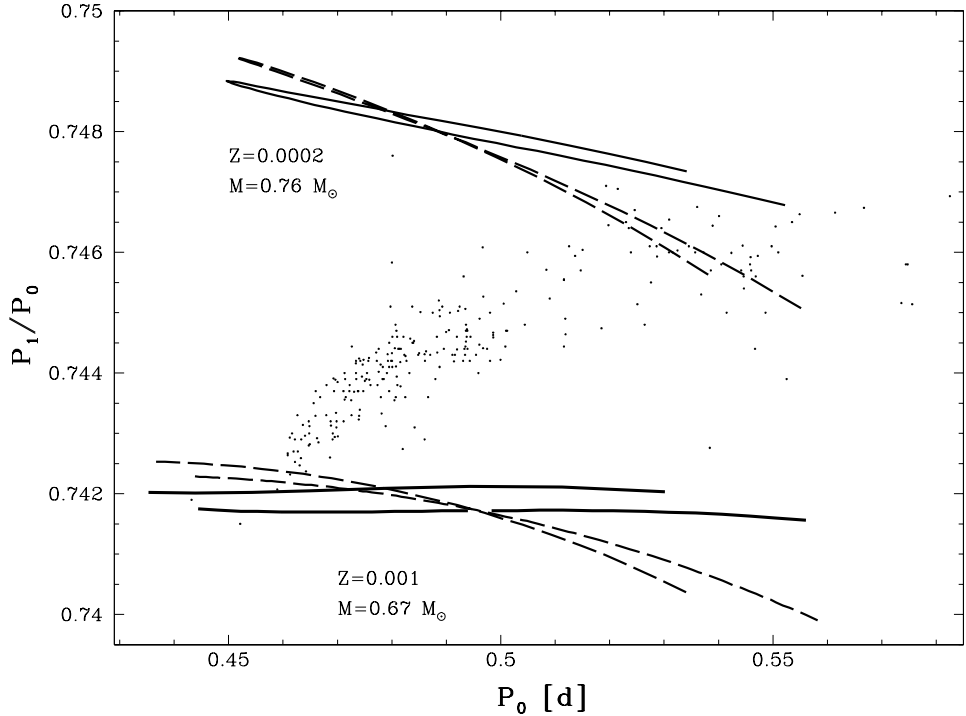,height=\linewidth,angle=270,bbllx=49,bblly=28,bburx=572,bbury=753}
 \caption[]{The effect of mixing length theory parameter, $\alpha$, on model
 position in the Petersen diagram. Results of pulsation calculations with
 $\alpha=1$ (solid lines) are compared with that calculated with $\alpha=2$
 (dashed lines). Dots correspond to RRd data.}
\label{f:alpha}
\end{figure}

%**************************************
\section{Summary}
%**************************************
\label{s:sum}

We have seen that the Petersen diagram for RRd stars in the Magellanic
Clouds may be explained by standard model calculations. The calculations
involved evolutionary models of horizontal branch stars and linear 
non-adiabatic calculations of radial pulsation periods. The agreement has
been achieved assuming metal abundances consistent with other determinations.

This successful explanation of the Petersen diagram may be regarded as a test
of our models. These models yield mean absolute magnitudes, 
$\left<M_V\right>\approx0.5$ mag. Hence we support the brighter luminosity 
scale for RR~Lyrae stars.

The range of metallicity needed to explain the whole extent of RRd band
is [Fe/H]$=(-2,-1.3)$ for both Magellanic Clouds. While in SMC the
objects appear to be uniformly distributed in this range, in LMC we see a
strong concentration in the range $(-1.7,-1.3)$.

We have discussed uncertainties in our inference on metallicities
following from the uncertainty of calculated values of period ratios.
Its primary sources are treatment of convection and nonlinear effects as
well as ambiguity of the heavy element abundance. All these effects
contribute to uncertainty of [Fe/H] on the level of $0.5$. All of them
should and can be reduced.

The observed width of the RRd band in the Petersen diagram may be explained
by the spread in masses at given metallicity. Petersen diagrams provide
a stringent constraint on RRd temperatures. As expected, these temperatures
correspond to the mid of the RR~Lyrae range. In detail however, there
is a difference between our RRd path in the HR diagram and that
found by nonlinear modeling. Our RRd strip is significantly more
inclined than the blue and red edges of the instability strip. The path
from nonlinear simulations has inclination more than 90 degrees.

%**************************************
\Acknow{We thank OGLE~Team for the unpublished data on RRd stars from
SMC and G{\'e}za~Kov{\'a}cs for discussions. We also acknowledge useful
comments of Giuseppe Bono. This work was supported in part by
the Polish State Committee for Scientific Research grant 2-P03D-14. One of
us (S.C.) was supported by MURST under the project ``Stellar evolution''.}

%**************************************

\end{document}